# Periodic Oscillations of Josephson-Vortex Flow Resistance in Oxygen-Deficient $Y_1Ba_2Cu_3O_x$

M. Nagao, S. Urayama, S. M. Kim, H. B. Wang, K. S. Yun, Y. Takano, T. Hatano, I. Iguchi, T. Yamashita, M. Tachiki and H. Maeda

National institute for Materials Science (NIMS), Tsukuba 305-0047, Japan,

M. Sato

Kitami Institute of Technology, 165 Koen-cho, Kitami 090-8507, Japan

*Abstract*— We measured the Josephson vortex flow resistance as a function of magnetic field applied parallel to the *ab*-planes using annealed $Y_1Ba_2Cu_3O_x$ intrinsic Josephson junctions having high anisotropy (~40) by oxygen content reduction. Periodic oscillations were observed in magnetic fields above 45-58 kOe, corresponding to dense-dilute boundary for Josephson vortex lattice. The observed period of oscillations, agrees well with the increase of one fluxon per two junctions ($H_p=\Phi_0/2Ls$), may correspond to formation of a triangular lattice of Josephson vortices as has been reported by Ooi *et al.* for highly anisotropic (≥200) Bi-2212 intrinsic Josephson junctions.

## I. INTRODUCTION

The crystalline structure of high-$T_c$ superconductors can be considered as a stack of Josephson coupled superconducting $CuO_2$ layers oriented perpendicularly to the *c*-axis. These junctions are called intrinsic Josephson junctions (IJJs)[1,2]. When a magnetic field is applied parallel to the *ab*-planes of the single crystal, Josephson vortices (JVs) carrying one flux quantum and whose cores are located in the non-superconducting layers, may form lattice structures[3]. When a bias current is applied along the *c*-axis, the JVs flow along the *ab*-direction because of the so-called Lorentz force and a flux-flow voltage, which is proportional to the flow velocity, appears.

The aim of this paper is to observe the JVs flow in $Y_1Ba_2Cu_3O_x$ (Y-123) IJJs, especially the periodic oscillations of the JVs flow resistance by the edge effect through which one could analyze the JVs lattice structure. So far as resistance oscillations in Y-123 IJJs, Ling *et al.* found oscillations in the dynamical resistance of Y-123 IJJs at the temperature near the midpoint of superconducting transition for mm-sized samples at the magnetic field 0-300 Oe[5]. At relatively high temperature (80.54 K), they observed peaks at $H\sim3\Phi_0/Ls=\Phi_0/L(s/3)$ for the series of samples with length 0.65-1.9 mm and explained the period by the junction thickness corresponding to the distance of $CuO_2$-Y-$CuO_2$ layers (0.34 nm≈$s$/3). Here $\Phi_0$, $L$ and $s$ express flux quantum, junction length perpendicular to the magnetic field and the period of junction array (=$c$-axis lattice constant of Y-123), respectively. The next peak was observed at ~$8\Phi_0/Ls$. The origin of these peaks remains open question. At slightly low temperature (79.80-79.50 K), they observed periodic oscillations in the dynamical resistance with period ranging from 11 to 16 Oe which was close to $\Delta H\sim\Phi_0/L(2s)$=13.5 Oe for the Y-123 IJJs with $L$=0.65 mm and offered several speculations for the doubling of $s$. The origin of the doubled period of junction array $2s$ has been depicted as the formation of the triangular lattice of the JVs by Ooi *et al.*[6], because the *c*-axis constant is doubled. They also found that the oscillation starts at above the certain magnetic field (~7 kOe). Since the oscillations found by Ling *et al.* starts from zero field with the period of 11 Oe (and then change to 16 Oe), there remains a question whether the origin of the oscillations in the two cases [Refs. 5 and 6] are identical or not.

Fabricating the artificial stacked array junctions, Krasnov *et al.* observed clear oscillations in the dynamical resistance of Nb-Cu (20/15 nm) supperlattice at the temperature near the superconducting transition (7.5 K) for 20 μm-ϕ sample at the magnetic field from 1.86 to 2.75 kOe with the period one fluxon per junction ($\Phi_0/Ls$). The period was explained by the periodic Fraunhofer dependence of Josephson critical current ($I_c(H)$) in junction array in the 3D regime where coherence length perpendicular to the layers is larger than the period of the array ($s$=35 nm).

It should be noted that the findings in $Bi_2Sr_2CaCu_2O_x$ (Bi-2212; ref-8) IJJs by Ooi *et al.* are for the flow resistance $R_{FF}=V_{FF}/I_{bias}$ at substantially low temperature compared to $T_c$ (from 4.2 K to ($T_c$-5) K) with precise periodicities and with high oscillation intensities[6]. Machida has studied the mechanism of these oscillations theoretically and elucidated the matching between the flowing triangular JVs lattice and the junction edges for the cause of the oscillations[9]. Koshelev has sketched out the distortion of the triangular JVs lattice at the edges and has pointed out that the origin of the oscillation derives from the inverse relation of the flow resistance to the $I_c(H)$ in junction array.

The configurations of the JVs lattice in stacked Josephson junctions have been discussed theoretically[10-13] and the magnetic field necessary for filling uniformly all the layers of the IJJs with JVs has been deduced as,

$$H_{d-d} = \alpha(\Phi_0/s\lambda_j), \qquad (1)$$

where $\gamma s$ expresses the size of the non-linear JV core along the layer (and frequently called as "the Josephson penetration depth $\lambda_j$ of IJJs"), $\gamma$ is the anisotropy factor and the $\alpha$ are $1/2\pi$, $1.4/2\pi$, $(1/12)^{1/2}$, and $(3/4)^{1/2}$, respectively depend on the theoretical approaches for Refs. 10-13. Experimentally, if JVs uniformly fill all the layers in the IJJs, each Josephson junction in the IJJs would show similar behavior as a function of the magnetic field. Then, the oscillation in JVs flow resistance appears because of the interference of JVs lattice with the edges; $H_{d-d}$ is a good measure of the starting field of the oscillations in JVs flow resistance $H_s$. In Ref. 6, Ooi *et al.* have suggested that the $H_s$ relates to the $H_{d-d}$ obtained in Ref. 11. Such fields are in inverse proportion to the anisotropy factor $\gamma$ from Eq. (1)[10-13]. The inverse relation has been proved by the decreased $H_s$ with increasing $\gamma$ observed in the Bi-based high-$T_c$ superconductors[14,15].

Note that at substantially low temperature compared to $T_c$, the oscillations in flux-flow resistance have been observed mostly in the Bi-based high-$T_c$ superconductors. Such oscillations should be observed generally in high-$T_c$ superconductors at low enough temperature suitable for applications, especially, in Y-123 phase. Because Y-123 has a high Josephson plasma frequency with a range beyond THz, which is significantly higher than that of Bi-2212 (~0.1THz)[16], the Y-123 IJJs are much more attractive candidates for actual THz device applications. However, the intrinsic Josephson effect observed in Y-123 was indecisive due to its low anisotropy which causes strong interlayer coupling[17]. In order to observe periodic oscillations of the JVs flow resistance at $T\ll T_c$, JVs are required to fill all layers in the IJJs, namely to form the lattice. The fully oxidized Y-123 ($\gamma<7$) requires extremely high magnetic fields, above 300 kOe, according to Eq. (1), because such fields are in inverse proportion to $\gamma$. To observe $H_{d-d}$ in Y-123, it is obviously necessary to increase the anisotropy of Y-123. Recently, we fabricated the Y-123 IJJs using high quality single-crystal whiskers and observed clear multi-branch structures which are closely comparable to that for Bi-2212[18,19] since we notice that the anisotropy $\gamma$ can be controllably increased with decreasing the oxygen content in the Y-123. In this paper, we report on the observation of the periodic oscillations of the JVs flow resistance in Y-123 IJJs with precise periodicities and with substantial oscillation intensities by increasing anisotropy by annealing the sample in vacuum. In addition, the starting field $H_s$ will be discussed in the relation with $\gamma$.

## II. EXPERIMENTAL

Y-123 single crystal whiskers were grown by the Te-doping method[19,20]. The whiskers used for the measurements had flat *ab*-plane surfaces about 25 μm in width and 2 mm in length. The thickness along the *c*-axis was about 5 μm. Four electrodes were placed on the Y-123 whisker using silver paste for transport measurements. We fabricated micron-scale *in-line-shaped* IJJs using a three-dimensional focused (Ga-)ion beam (FIB) etching method[21], eliminating the non-uniformity, which is a typical problem in the surface-junction of mesa-type IJJs. We estimated the size of the IJJs from a scanning ion microscope (SIM) image; the lengths of the IJJs perpendicular to the magnetic field (*L*) were 1.0, 1.9 and 4.3 μm for IJJs-1, -2 and –3, respectively. The depths along the field direction within the *ab*-planes were about 5 μm for the three IJJs. The relatively long scales along the field direction were chosen so as to enhance the edge effect. The total thickness *t* of the IJJs was about 0.2-0.4 μm along the *c*-axis, corresponding to approximately 170-340 junctions in the stack. Table I summarizes the parameters of Y-123 IJJs examined in this paper. Figure 1 (a) shows a SIM image of the IJJ-2 fabricated into the in-line-shape. After the FIB-etching, the IJJs were annealed at 400 °C in vacuum for 1-2 hours for oxygen reduction in order to increase their anisotropies.

The transport properties of IJJs were measured by "the physical property measurement system" (Quantum Design, PPMS) equipped with a 70 kOe split magnet and a rotational sample stage. Prior to the measurements, the *ab*-plane of the IJJs was precisely adjusted to the magnetic field by the angular dependence of the flow resistance (*lock-in* curve) under the constant magnetic field at a temperature just below $T_c$. When the field direction is close to parallel to the *ab*-plane, the vortices become a *lock-in* state[22,23], which is expected to be free from pancake vortex which goes across the superconducting $CuO_2$ planes. The flux-flow voltages were measured by the standard four-probe method with a dc current under the increasing magnetic field *H* parallel to the *ab*-planes (*H*//*ab*-plane). We define the flow resistance as this voltage divided by the dc bias current for the measurement.

| Sample | IJJs-1 | IJJs-2 | IJJs-3 |
|---|---|---|---|
| $L$ (μm) | 1.0 | 1.9 | 4.3 |
| $D$ (μm) | 4.9 | 4.95 | 5.1 |
| $t$ (μm) | 0.4 | 0.2 | 0.3 |
| $T_c$ (K) | 37 | 22.5 | 35 |
| $\gamma$ (estimated) | 37 | 46 | 39 |
| $H_p$ calc.(kOe) | 8.8 | 4.7 | 2.1 |
| $H_p$ obs.(kOe) | 8.6 | 5.5 | 2.0 |
| $H_{d-d}$ calc.(kOe) | 65 | 52 | 63 |
| $H_{d-d}$ obs.(kOe) | 56 | 45 | 60 |

Table I. The parameters of Y-123 IJJs samples used for the flux-flow resistance oscillations measurements.

## III. RESULTS AND DISCUSSION

Figure 1 (b) shows typical current-voltage (*I-V*) characteristics of the Y-123 IJJs under the zero field at 5 K [19]. The *I-V* characteristics show clear multiple branches with large hysteresis typical for IJJs, suggesting that our *in-line-shaped* Y-123 IJJs made of single crystal whiskers are homogeneous and of good quality. A weak junction, which had lower $I_c$ than other IJJs, was not found.

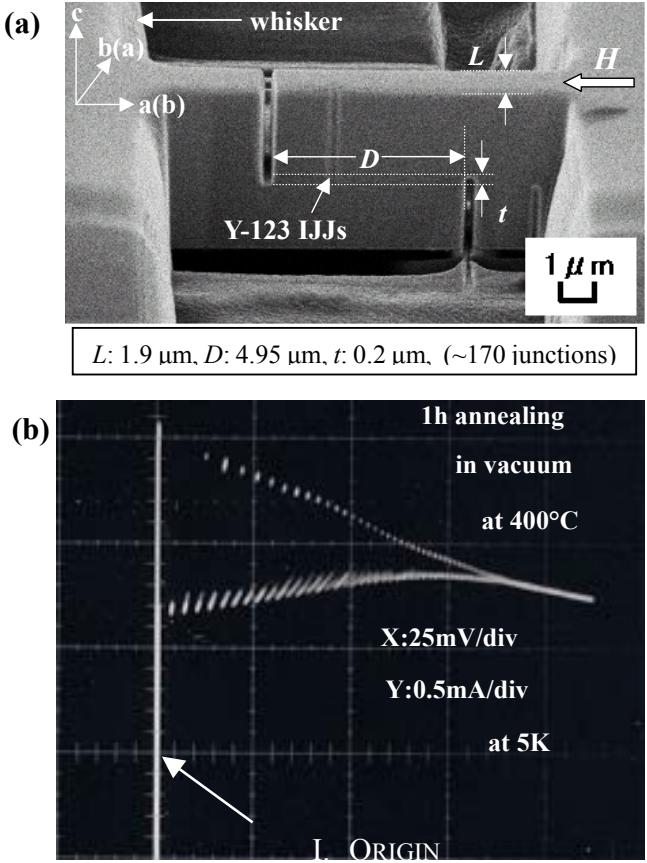

Fig. 1 (a). Scanning ion microscope (SIM) image of the in-line-shaped Y-123 intrinsic Josephson junctions. The cross-sectional area and thickness of junctions are about 9.4 μm² and 0.2 μm, respectively. (b). Typical current-voltage (*I-V*) characteristics of the Y-123 intrinsic Josephson junctions under the zero field at 5 K (See Ref. 19).

Three oxygen reduced whiskers, fabricated in bridge-shape, were prepared with different oxygen contents for estimating the anisotropy of the whisker whose $T_c$s range 40-90 K. Although this method is introduced in Refs. 24 and 25, we describe it in brief for clarity. We measured the angular dependence of the resistivity $\rho$ at various magnetic fields $H$ in a flux liquid state. We evaluated a reduced field by the effective mass model,

$$H_{red}=H(sin^2\theta+\gamma^{-2}cos^2\theta)^{1/2}, \quad (2)$$

where $\theta$ is the angle between the *ab*-plane and the magnetic field [26]. We plotted the $\rho$-$H_{red}$ relation, and estimated the anisotropic factor $\gamma$, which gives the best scaling for the $\rho$-$H_{red}$ relations. Figure 2 (a) shows three data points, from which the critical temperature ($T_c$) dependence of anisotropy ($\gamma$) ($\gamma$-$T_c$ relation) can be fitted by a straight line. Empirically, the anisotropy increases with decreasing the oxygen content in the Y-123.

It is difficult to obtain homogeneous oxygen-deficient samples with $T_c$s lower than 40 K by annealing the bridge-shaped samples. Thus, in order to obtain IJJs with $T_c$<40 K, the in-line-shaped IJJs of Y-123 are annealed in vacuum. Figure 2 (b) shows resistivity-temperature ($\rho$-$T$) characteristics of the IJJs-2 with a dc current of 10 μA. Zero resistance is obtained at 22.5 K, and the anisotropy of the sample is estimated as $\gamma$~46 (solid square in Fig. 2 (a)) from linear extrapolation of the obtained $\gamma$-$T_c$ relation. In the similar manner, the anisotropies of IJJ-1 and IJJ-3 are expected as $\gamma$~37 and $\gamma$~39, respectively.

It should be noted, Rapp *et al.* reported that $\gamma$=83 for the polycrystalline thin film with $T_c$=42 K[17] which is factor 2-3 larger than ours. It should be emphasized that our estimation for $\gamma$ has been done using single crystals at the superconducting state.

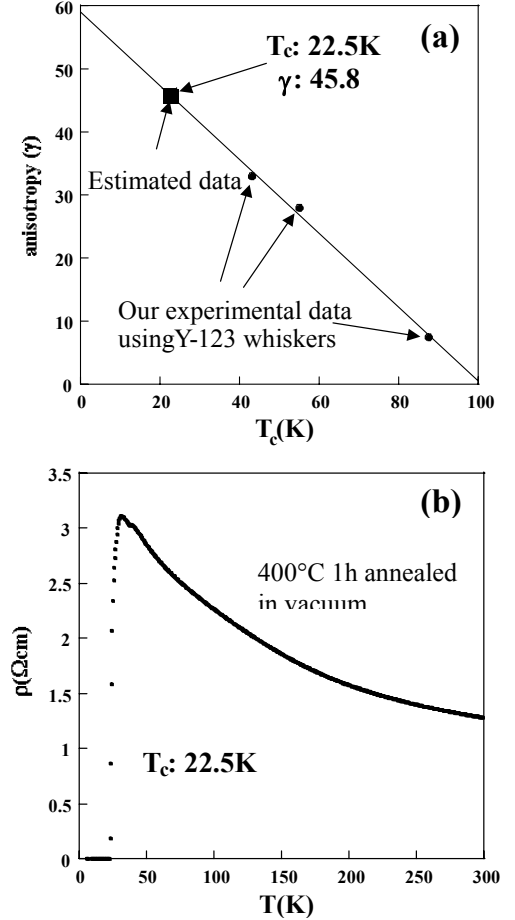

Fig. 2 (a). The critical temperature ($T_c$) dependence of anisotropy ($\gamma$) for Y-123 single crystal whiskers. 2 (b). Resistivity-temperature ($\rho$-$T$) characteristics of the annealed Y-123 intrinsic Josephson junctions. The Y-123 intrinsic Josephson junctions are annealed at 400 °C for one hour in vacuum.

Figure 3 shows the JVs flow resistance of the IJJs as a function of the magnetic field (*R-H* curve) with dc currents of (a) 140-180µA, (b) 3-15 µA and (c) 30-100 µA for IJJ-1, IJJ-2 and IJJs-3 respectively at 5 K. In the field higher than 56 kOe, 45 kOe and 60 kOe for IJJ-1, IJJ-2 and IJJs-3 respectively, the flow resistance shows periodic oscillations with period $H_p \sim \Phi_0/2Ls$. The observed onset fields of the periodic oscillations in JVs flow resistance shows good agreement with the dilute-dense boundary predicted by Koshelev in Eq. (1) with $\alpha=1/2\pi$ [10], which are $H_{d-d}=\Phi_0/2\pi\gamma s^2$=65 kOe, 52 kOe and 63 kOe, for IJJ-1, IJJ-2 and IJJs-3 respectively.

In the *R-H* curve for IJJs-2 shown in Fig. 3 (b), the maxima of oscillations increase with increasing the magnetic field. The periodic oscillations are unclear in the dc bias current lower than 5 µA, mainly due to the noise level of our measurement system. On the other hand, the oscillation in flux-flow resistance disappears above 14 µA. This current level is 4.5 % of the critical current (310 µA) at the zero field. The clear periodic oscillations appear in the current level between 6 and 12 µA.

The observed periods of flow resistance oscillations $H_p$ are approximately 8.6, 5.5 and 2.0 kOe, for IJJs-1, IJJs-2 and IJJs–3, respectively. According to the equation $H_p=\Phi_0/2Ls$, and the measured *L* from each SIM image, the calculated periods are 8.8, 4.7 and 2.1 kOe, respectively. The observed periods show good agreement with $H_p=\Phi_0/2Ls$, that is, one fluxon per two junctions. The formation of the triangular JVs lattice could be the origin of this period as has been proposed by Ooi *et al*. for the Bi-2212 IJJs[6,27].

The flow resistance oscillations are hardly observed at 10 K. Only from the IJJs-3, we succeeded to observe oscillations as shown in Fig. 4 at the low current levels. Although the oscillation intensity is small, the phase and the period of oscillations are identical to those observed at 5 K shown in Fig. 3 (c).

Recently, Ustinov and Pedersen have proposed that both half and one fluxon period oscillations can be explained by the Fiske-steps of a single Josephson junction[28]. As has been shown in the uniform multibranch structure in our IJJs, it is not likely that only one junction contributes the flow voltage. Nevertheless, it is necessary to check whether the Fiske-step from the weakest single-layer in the IJJs could be an alternative origin of JVs flow resistance oscillations or not. We estimate the Fiske-step voltage of a single-layer junction. The Fiske-step voltage is expressed as

$V_f=(\Phi_0/2L)c_s$   (3)

where $c_s$ is the Swihart velocity [$c_s=c_o(d_I/\varepsilon_r d)^{1/2}$], $c_o$ is the velocity of light in vacuum, $\varepsilon_r$(=10 for Y-123) is the dielectric constant, $d \sim d_I + 2\lambda_L$, $\lambda_L$(=140 nm for Y-123) is the London penetration depth and $d_I$ (=0.84 nm for Y-123) is the thickness of non-superconducting layer[29-31]. This Fiske-step voltage is estimated as $V_f \sim 2.8$ mV for IJJ-2 having $L=1.9$ µm. It suggests that the JVs flow voltage in our IJJs-2 during the flux-flow resistance measurement (Shown in Fig.3) are a few order lower than that of Fiske step voltage in a single-layer junction, therefore the periodic oscillations observed in this experiment is not due to the Fiske steps.

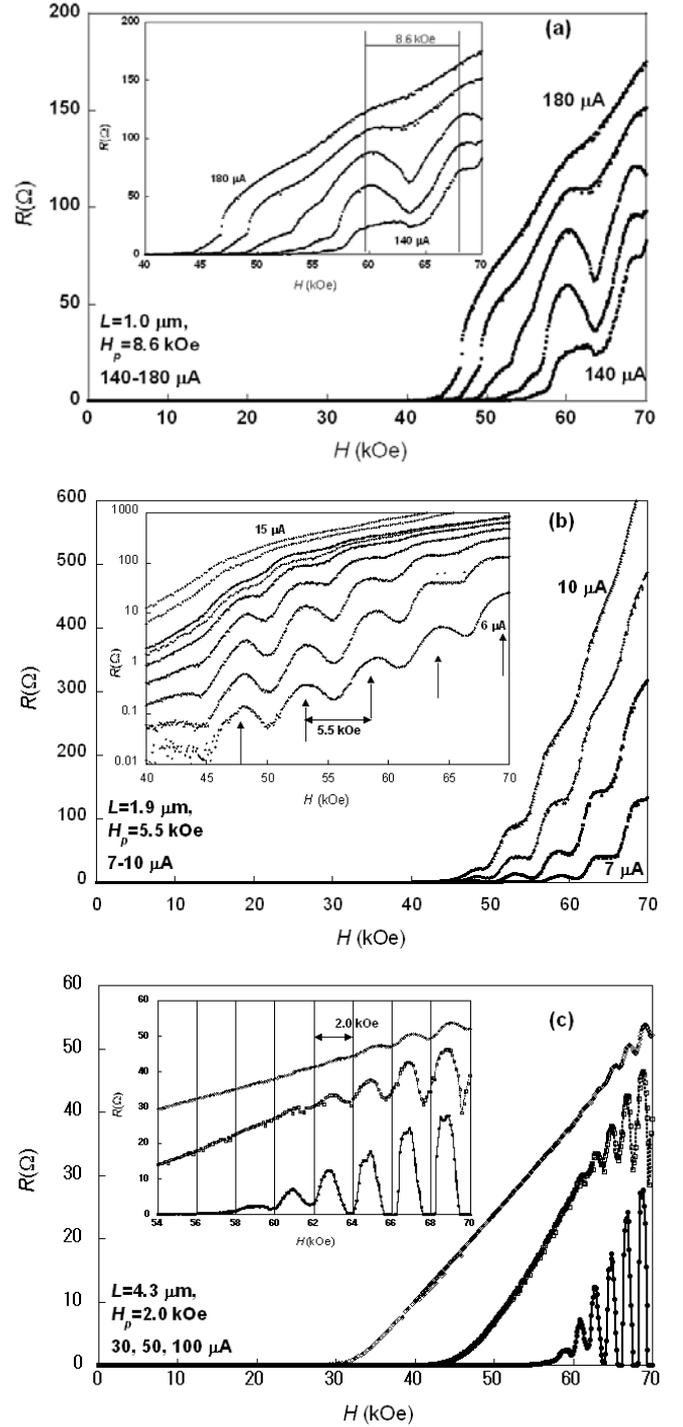

**Fig. 3 Josephson vortices (JVs) flow resistance as a function of the magnetic field parallel to the *ab*-planes at 5 K. Shown in the inset is the close-up of the flow resistance oscillations. (a) IJJs-1 with *L*=1.0 µm, (b) IJJ-2 with *L*=1.9 µm, and IJJ-3 with *L*=4.3µm.**

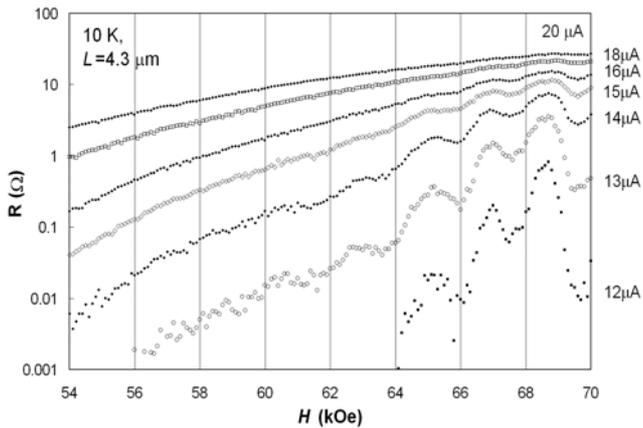

**Fig. 4 Josephson vortices (JVs) flow resistance as a function of the magnetic field parallel to the *ab*-planes at 10 K for IJJ-3 with *L*=4.3μm.**

IV. CONCLUSION

We have successfully observed the periodic oscillations of vortex flow resistance in oxygen deficient Y-123 intrinsic Josephson junctions which have anisotropic factors ~40. The period of flux-flow oscillations showed good agreement with $\Phi_0/2Ls$, indicating that "one" Josephson vortex is added per "two" intrinsic Josephson junctions for each oscillation period. However, such periodic oscillation only appears above the magnetic fields of 45-58 kOe, which might correspond to the boundary between dilute and dense Josephson vortex lattices. Obviously, this threshold field is higher than those observed in high anisotropic Bi-2212 intrinsic Josephson junctions because of the one order low anisotropies for the oxygen-reduced Y-123 IJJs studied here.

ACKNOWLEDGMENT

The authors would like to thank Dr. X. Hu of the National Institute for Materials Science for useful discussions, and Ms. R. Yamashita of the National Institute for Materials Science for critical reading of the manuscript.